\documentclass[12pt]{article}
\usepackage{amssymb}
\usepackage{amsmath}
\usepackage{amsfonts}
\begin{document}

\baselineskip=24pt

\bibliographystyle{unsrt}
\vbox {\vspace{6mm}}
\bigskip
\bigskip
\bigskip
\bigskip
\begin{center} {\bf Tomographic probability representation for quantum fermion fields}
\end{center}

\begin{center}
{\bf V. A. Andreev$^1$, M. A. Man'ko$^1$, V. I. Man'ko$^1$, Nguyen
Hung Son$^2$, Nguyen Cong Thanh$^2$, Yu.P.Timofeev$^1$ and
S.D.Zakharov$^1$}
\end{center}

\begin{center}$^1$ {\it P. N. Lebedev Physical Institute\\ Leninskii Prospect 53, Moscow
119991, Russia\\} $^2$ {\it Institute of Physics\\ 10 Dao Tan
Street, Ha Noi, Viet Nam\\}
\end{center}

\bigskip
\bigskip
\bigskip

\begin{abstract}
\noindent Tomographic probability representation is introduced for fermion
fields. The states of the fermions are mapped onto probability
distribution of discrete random variables (spin projections). The
operators acting on the fermion states are described by fermionic
tomographic symbols. The product of the operators acting on the
fermion states is mapped onto star-product of the fermionic
symbols. The kernel of the star-product is obtained. The
antisymmetry of the fermion states is formulated as the specific
symmetry property of the tomographic joint probability
distribution associated with the states.
\end{abstract}

\section{Introduction}

\indent In quantum mechanics the states are described (in standard
formulation) by vectors in Hilbert space \cite{Dirac} (wave
function \cite{Scr}) or density operators \cite{Landau}, \cite{van
Neumann}. In the quantum field theory one considers not only
states of the fixed numbers of particles but also the process of
creation and annihilation of the particles (see \cite{Bog},
\cite{NVH}). The formulation of quantum field theory is based on
the second quantization procedure (see, e.g \cite{Berezin}). The
fields are divided into two classes: boson fields and fermion
fields. The states of the boson fields must be symmetric with
respect to permutation of the particles (bosons). But the states
of the fermion fields must be antisymmetric ones. Recently
probability representation of quantum states (tomographic
probability representation) in quantum mechanics was introduced
\cite{Manc 1996}, (see also recent review \cite{Vent}). In quantum field theory the tomographic
probability representation was studied  for bosons still now in
few works \cite{Rosa}, \cite{Stern1}, \cite{Stern2}. The more systematic
tomographic approach to study quantum boson fields (scalar fields)
was suggested in our previous work \cite{Son}. In this work the
star-product formalism for tomographic probabilities \cite{Mar} describing
the boson field states was presented. The aim of our work is to
suggest the tomographic probability representation for fermion
fields. We consider the identical spin-1/2 particles and their
states in framework of free fields. To do this we review and
apply spin-tomographic description of fermion states in quantum
mechanics \cite{D},\cite{O},\cite{An}. Then we extend the
spin-tomographic approach to the fermion fields and their states
description. It is worthy to note that the explicit matrix
representation of fermion anticommutation relation was discussed
in \cite{Dav}. In our construction we use and develop the approach
given in this book.

The paper is organized as follows. In Sec. 2 we present the
spin-tomographic scheme for one and two particles states. In Sec.
3 the star-product formalism is developed for the spin-tomographic
symbols of the operators acting in the space of fermion states. In
Sec. 4 matrix representation of fermionic operators is given. In
Sec.5  the important operators like creation and annihilation
operators as well as vacuum state density matrix will be studied
in the tomographic framework. The conclusion and perspective are
given in Sec. 6

\section{State of spin-1/2 particle}

First we study the pure state of spin-1/2 particle in quantum
mechanics. The state is described by a spinor
\begin{equation}\label{A2}
|\Psi\rangle=\begin{pmatrix}a \\ b\end{pmatrix},
\end{equation}

\noindent where $a, b$ are complex numbers and normalization
condition holds
\begin{equation}\label{B2}
|a|^2 + |b|^2 = 1.
\end{equation}

The tomogram of the state (\ref{A2}) is determined by the
probability vector $\vec {\omega}$ with the two component
${\omega(+\frac{1}{2},u)}, {\omega(-\frac{1}{2},u)}$ where
\begin{eqnarray}\label{C2}
\begin{aligned}
 \omega(m,u) & = {\mid {\langle m}\mid u \mid \Psi \rangle
\mid}^2, \\
m & = \pm \frac{1}{2},
\end{aligned}
\end{eqnarray}

\noindent and the unitary matrix $u$ rotating the spinor
(\ref{A2}) in terms of Euler angels $\varphi$, $\theta$ and $\psi$
is
\begin{equation}\label{D2}
u=\begin{pmatrix}
  \cos\frac{\theta}{2} e^{\frac{\iota(\varphi+\psi)}{2}} & \sin\frac{\theta}{2} e^{\frac{\iota(\varphi-\psi)}{2}} \\
  -\sin\frac{\theta}{2} e^{\frac{-\iota(\varphi-\psi)}{2}} & \cos\frac{\theta}{2} e^{\frac{-\iota(\varphi+\psi)}{2}}
\end{pmatrix}
\end{equation}

For mixed state with density matrix
\begin{equation}\label{E2}
\rho=\left(%
\begin{array}{cc}
  \rho_{11} & \rho_{12} \\
  \rho_{21} & \rho_{22} \\
\end{array}%
\right),
\end{equation}

\noindent the tomogram is given by diagonal elements of rotated
density matrix
\begin{equation}\label{F2}
\omega (m,u) = (u{\rho}{u^{+}})_{mm}.
\end{equation}

As was sown in \cite{D},\cite{O} the tomogram determines the
density matrix. Also the tomogram satisfies the normalization
condition
\begin{equation}\label{G2}
\sum\limits_{m=-1/2}^{1/2} \omega (m,u) = 1,
\end{equation}

\noindent for arbitrary values of Euler angles. For two particles
with spin 1/2 the tomogram is introduced as joint probability
distribution of two random spin projection $m_1$, $m_2$
\begin{equation}\label{H2}
\omega (m_{1}, m_{2}, u) =
(u{\rho(1,2)}{u^{+}})_{m_1m_2,m_1m_2}\,.
\end{equation}

Here $\rho(1,2)$ is density $4{\times}4$ matrix of the two particle
spin state. The indices $m_1, m_2 = \pm {\frac12}$ and
correspond to pure spin state $\mid m_1m_2\rangle$ basis. Thus
\begin{equation}\label{I2}
\rho(1,2)_ {m_1m_2,n_1n_2}= \langle m_1m_2\mid{\hat\rho(1,2)} \mid
n_1n_2\rangle.
\end{equation}

For pure state of two particles $|\Psi\rangle$ the tomogram
according to formulaes (\ref{H2}), (\ref{I2}) reads
\begin{equation}\label{J2}
\omega(m_1,m_2,u) =  {\lvert {\langle {\Psi}\mid u \mid
m_1m_2\rangle} \rvert}^2.
\end{equation}

The unitary $4{\times}4$-matrix $u$ is tensor product of two $2{\times}
2$-matrices (\ref{D2}) each depending on their Euler angles
\begin{equation}\label{K2}
u = u_1\bigotimes u_2.
\end{equation}

It is clear that for multiparticle state construction of
tomographic probability distribution (join probability
distribution) is straightforward. The tomograms (\ref{H2}), (\ref{J2}) are
nonnegative and satisfiy the normalization condition
\begin{equation}\label{L2}
\sum_{m_1=-1/2}^{1/2}\sum_{m_2=-1/2}^{1/2}
\omega(m_1,m_2,u) =1,
\end{equation}

\noindent for arbitrary Euler angles determining matrices $u_1$
and $u_2$.

\section{Star-product for spin-1/2 operators}

The quantum operators can be mapped onto the functions
\cite{Berezin}, \cite{Stratonov}, \cite{Fronsdal}. The product of
the functions called symbols of the operators is the star-product
which is associative but not commutative one. The product is
determined by a kernel. For standard pointwise product of
functions the kernel is local one. In general case the product
kernel is non-local. The scheme of the star-product construction
is the following \cite{Mar}. If one has an operator $\hat A$ and specific
pair of operators $\hat U(\vec x)$ called dequantizer and $\hat
D(\vec x)$ called quantizer satisfying relation
\begin{equation}\label{A4}
Tr\hat D(\vec x) \hat U(\vec x') = \delta (\vec x - \vec x'),
\end{equation}

\noindent the symbol of operator $\hat A$ is defined as
\begin{equation}\label{B4}
f_A(\vec x) = Tr\hat A \hat U(\vec x).
\end{equation}

The operator is reconstructed from its symbol by means of the
quantizer
\begin{equation}\label{C4}
\hat A = \int f_A(\vec x)\hat D(\vec x)d{\vec x}.
\end{equation}

In (\ref{A4})-(\ref{C4}) the coordinates $\vec x$ can be any set
of continuous or discrete variables $\vec x = (x_1, x_2,…, x_n)$
and integration in (\ref{C4}) is understood as integration over
continuous variables and summation over the discrete variables.
The symbols of the operators $f_A(\vec x)$ and $f_B(\vec x)$ provide
the symbol of operator $f_C(\vec x)$ if
\begin{equation}\label{D4}
\hat C = \hat A \hat B, \end{equation}

\noindent using the star-product definition
\begin{equation}\label{E4}
f_C(\vec x): = f_{AB}(\vec x) = Tr \hat A \hat B \hat U(\vec x)\,.
\end{equation}

The symbol $f_C(\vec x)$ can be presented in the integral form
\begin{equation}\label{F4}
f_C(\vec x) = \int f_A(\vec y)f_B(\vec z)K(\vec y, \vec z, \vec x)
d{\vec y}d{\vec z}.
\end{equation}

Here non-local kernel determining the star-product of symbols
reads
\begin{equation}\label{G4}
K(\vec y, \vec z, \vec x) = Tr (\hat D(\vec y) \hat D(\vec
z) \hat U(\vec x).
\end{equation}

For spin-1/2 particle we present the tomographic star-product of
the spin-tomograms of the operators. The coordinate $\vec x = (m,
\theta, \psi) \equiv (m, \vec n)$ contains three variables. The
discrete variable $m = \pm \frac{1}{2}$ is the spin projection on
z-axis. The continuous variables $\theta$ and $\varphi$ are
coordinates of a point on the sphere $0 \le \psi \le 2\pi, 0 \le
\theta \le \pi$. These coordinates can be described by unit vector
$\vec n = (sin{\theta}cos{\psi}, sin{\theta}sin{\psi},
cos{\theta})$, ${\vec n}^2 = 1$.

The dequantizer $\hat U(\vec x) = \hat U(m, \vec n)$ which is $2{\times}2$
matrix reads \cite{Fil}
\begin{eqnarray}\label{H4}
\hat U(m, \vec n) = \frac{1}{2} \begin{pmatrix} 1 & 0\\0 & 1
\end{pmatrix}
+ m \begin{pmatrix}  \cos{\theta} & - e^{\iota \psi}\sin{\theta} \\
  - e^{-\iota \psi}\sin{\theta} & -\cos{\theta}
\end{pmatrix},
\end{eqnarray}

\noindent and the quantizer $\hat D(\vec x) = \hat D(m, \vec n)$
is
\begin{eqnarray}\label{I4}
\hat D(m, \vec n) = \frac{1}{2}
\left(%
\begin{array}{cc}
1 & 0 \\ 0 & 1
\end{array}%
\right)
+ 3m \left(%
\begin{array}{cc}
\cos{\theta} & - e^{\iota \psi}\sin{\theta} \\
  - e^{-\iota \psi}\sin{\theta} & -\cos{\theta} \\
\end{array}%
\right).
\end{eqnarray}

The symbol of arbitrary operator is defined by equality
\begin{equation}\label{J4}
f_A(m,\vec n) = Tr\hat A \hat U(m,\vec n),
\end{equation}

\noindent where the matrix of the operator is
\begin{equation}\label{K4}
A=\left(%
\begin{array}{cc}
  A_{11} & A_{12} \\
  A_{21} & A_{22} \\
\end{array}%
\right),
\end{equation}

\noindent i.e. the symbol reads
\begin{equation}\label{L4}
f_A(m,\vec n) = \frac{1}{2} + m [A_{11}\cos\theta -
e^{-\iota\psi}\sin\theta A_{12} + A_{21}(-
e^{\iota\psi}\sin\theta) - A_{22}\cos\theta].
\end{equation}

\noindent
One can check that condition (\ref{A4}) is given by symbolic formula
\begin{equation}\label{M4}
\delta (\vec x - \vec x') = Tr\hat D(m_1, {\vec n}_1) \hat U(m_2,
{\vec n}_2) = \frac{1}{2} + 6m_1m_2({\vec n}_1{\vec n}_2).
\end{equation}

The symbol with delta-function in Eq.(\ref{M4}) is not usual Dirac delta
function (or Kronecker one). It only means that the right-hand side of this equation (\ref{M4})
is equivalent to matrix element of identity operator acting in the set of tomograms.

\vspace{5mm}

The tomogram (\ref{C2}) of pure quantum state $|\Psi\rangle$ is defined in framework of
presented star-product scheme using the density operator $ \hat{\rho} = |\Psi\rangle\langle\Psi|$ and
corresponding density matrix in (\ref{F2}).

The kernel of star-product reads \cite{Fil}
\begin{eqnarray}\label{N4}
\begin{aligned}
K(m_1\vec {n_1},m_2\vec {n_2},m_3\vec {n_3}) & = Tr(\hat
D(m_1,\vec {n_1})\hat D(m_2,\vec {n_2})\hat U(m_3,\vec {n_3}))\\
& = \frac{1}{4} + 3m_1m_2(\vec{n_1}\vec{n_2}) + 9m_1m_3
(\vec{n_1}\vec{n_3}) + 9m_2m_3(\vec{n_2}\vec{n_3}) \\ &
\hspace{5mm} + 18\iota m_1m_2 m_3(\vec{n_1}\vec{n_2}\vec{n_3})
.
\end{aligned}
\end{eqnarray}

\section{Matrix representation of fermion operators.}
Let us construct matrix representation of the creation and annihilation
fermion operators for N fermions. If $N=1$ one can consruct the $2{\times}2$
matrices
\begin{equation}\label{A5}
a_1=\left(%
\begin{array}{cc}
  0 & 1 \\
  0 & 0 \\
\end{array}%
\right),\hspace{10mm}
{a_1}^{+}=\left(%
\begin{array}{cc}
  0 & 0 \\
  1 & 0 \\
\end{array}%
\right),
\end{equation}

\noindent which satisfy the relations
\begin{equation}\label{B5}
\begin{aligned}
a_1a_1 + a_1a_1 & = 0,\\
{a_1}^{+}{a_1}^{+} + {a_1}^{+}{a_1}^{+} & = 0,\\
a_1{a_1}^{+} + {a_1}^{+}a_1 & = 1,
\end{aligned}
\end{equation}

\noindent which can be rewritten as
\begin{eqnarray}\label{C5}
\begin{aligned}
a_ia_j + a_ja_i & = \{a_i,a_j\}= 0,\\
{a_i}^{+}{a_j}^{+} + {a_j}^{+}{a_i}^{+} & = \{{a_i}^{+},{a_j}^{+}\} = 0,\\
a_i{a_j}^{+} + {a_j}^{+}a_i & = \{a_i,{a_j}^{+}\} = \delta_{ij},
\end{aligned}
\end{eqnarray}

\noindent where $i,j = 1$.
If $i,j = 1,2$ the relation (\ref{C5}) can be satisfied by $4{\times}4$ matrices
\begin{equation}\label{D5}
\begin{aligned}
a_1 =
\begin{pmatrix}
1 & 0\\
0 & 1
\end{pmatrix}
 \otimes
\begin{pmatrix}
0 & 1\\
0 & 0
\end{pmatrix}, & \hspace{10mm}
{a_1}^{+} =
\begin{pmatrix}
1 & 0\\
0 & 1
\end{pmatrix}
\otimes
\begin{pmatrix}
0 & 0\\
1 & 0
\end{pmatrix} \\
a_2=
\begin{pmatrix}
0 & 1\\
0 & 0
\end{pmatrix}
\otimes
\begin{pmatrix}
1 & 0\\
0 & -1
\end{pmatrix},& \hspace{10mm}
{a_2}^{+} =
\begin{pmatrix}
0 & 0\\
1 & 0
\end{pmatrix}
\otimes
\begin{pmatrix}
1 & 0\\
0 & -1
\end{pmatrix}.
\end{aligned}
\end{equation}

If $i,j = 1,2,3$ the relations (\ref{C5}) are satisfied by the $8{\times}8$-matrices which we present in the form of tensor products
\begin{equation}\label{E5}
\begin{aligned}
a_1 & = 1 \otimes 1 \otimes \sigma_{+}, \\
a_2 & = 1 \otimes \sigma_{+} \otimes \sigma_{z}, \\
a_3 & = \sigma_{+} \otimes \sigma_{z} \otimes \sigma_{z}.
\end{aligned}
\end{equation}

\noindent and corresponding matrices ${a_j}^{+}$ associated with fermion creation operators.

For $N=4$ i.e. $i,j = 1,2,3,4$ in (\ref{C5}) the construction of the matrices $a_j$ looks as follows
\begin{equation}\label{F5}
\begin{aligned}
a_1 & = 1 \otimes 1 \otimes 1 \otimes \sigma_{+}, \\
a_2 & = 1 \otimes 1 \otimes \sigma_{+} \otimes \sigma_{z}, \\
a_3 & = 1 \otimes \sigma_{+} \otimes \sigma_{z} \otimes \sigma_{z}, \\
a_4 & = \sigma_{+} \otimes \sigma_{z} \otimes \sigma_{z} \otimes \sigma_{z}.
\end{aligned}
\end{equation}

We introduced standard notations for $2{\times}2$ matrices
\begin{eqnarray}\label{G5}
\begin{aligned}
1 = \left(%
\begin{array}{cc}
  1 & 0 \\
  0 & 1 \\
\end{array}%
\right),
\sigma_{+} = \left(%
\begin{array}{cc}
  0 & 1 \\
  0 & 0 \\
\end{array}%
\right),
\sigma_{z} = \left(%
\begin{array}{cc}
  1 & 0 \\
  0 & -1 \\
\end{array}%
\right),
\sigma_{-} = (\sigma_{+})^{+} = \left(%
\begin{array}{cc}
  0 & 0 \\
  1 & 0 \\
\end{array}%
\right).
\end{aligned}
\end{eqnarray}

The generalization of the construction for arbitrary number $N$ is
straightforward, i.e.
\begin{eqnarray}\label{H5}
\begin{aligned}
a_i = 1^{(1)} \otimes \cdots \otimes 1^{(N-i)} \otimes {\sigma_{+}}^{(N-i+1)} \otimes
{\sigma_{z}}^{(N-i+2)} \cdots \otimes {\sigma_{z}}^{(N)}.
\end{aligned}
\end{eqnarray}

Corresponding matrices ${a_i}^{+}$ are obtained by changing in (\ref{F5}) all $\sigma_{+}$ by $\sigma_{-}$ matrices.

The vector $\mid vac\rangle$ which satisfies the condition
\begin{equation}\label{I5}
a_j \mid vac\rangle = 0, j=1,2,\dots N,
\end{equation}

\noindent has the form of tensor product
\begin{equation}\label{J5}
\mid vac\rangle = \begin{pmatrix}1 \\ 0\end{pmatrix}
\otimes \begin{pmatrix}1 \\ 0\end{pmatrix}
\otimes \cdots \otimes \begin{pmatrix}1 \\ 0\end{pmatrix}.
\end{equation}

For $N=2$
\begin{equation}\label{K5}
\mid vac\rangle = \begin{pmatrix}1 \\ 0 \\ 0 \\ 0\end{pmatrix}.
\end{equation}

The density matrix of this state reads
$
\left(%
\begin{array}{cccc}
  1 & 0 & 0 & 0 \\
  0 & 0 & 0 & 0 \\
  0 & 0 & 0 & 0 \\
  0 & 0 & 0 & 0 \\
\end{array}
\right)$.

The states
\begin{equation}\label{M5}
\mid \dots i \dots j \dots \rangle =
{a_i}^{+}{a_j}^{+} \mid vac\rangle,
\end{equation}

\noindent are antisymmetric
\begin{equation}\label{N5}
\mid \cdots i \cdots j \cdots \rangle =
- \mid \cdots j \cdots i \cdots \rangle,
\end{equation}

\noindent by construction

\section {Tomographic representation of Fermi-operators.}

Let us express the creation and annihilation Fermi-operators in
tomographic form. To do this we first express the separate factors
ingredient in the tomographic form. There are three such factors,
namely factors given by (\ref{G5}).

According to general construction the spin-tomographic symbols
of the matrix factors are calculated using the diagonal matrix
elements of the matrices $u\sigma_{+}u^{+}$, $u\sigma_{z}u^{+}$,
$u 1 u^{+}$ as well as $u\sigma_{-}u^{+}$. We denote three symbols
$\Omega_{+}(m, \vec n)$, $\Omega_{z}(m, \vec n)$, $\Omega_{1}(m, \vec n)$
and $\Omega_{-}(m, \vec n)$ respectively. By construction
\begin{equation}\label{C6}
\Omega_{+}(m, \vec n) = {\Omega_{-}(m, \vec n)}^{*}.
\end{equation}

We remind that $m$ takes two values $m=\pm \frac{1}{2}$ and the unit vector
$\vec n = (sin{\theta}cos{\psi}, sin{\theta}sin{\psi},cos{\theta})$.

The symbol of identity operator reads
\begin{equation}\label{D6}
\Omega_{1}(+\frac{1}{2}, \vec n) =
\Omega_{1}(-\frac{1}{2}, \vec n) = 1.
\end{equation}

One can check that the symbol of the operators $\sigma_{z}$ has the following
values
\begin{equation}\label{E6}
\Omega_{z}(+\frac{1}{2}, \vec n) = cos\theta,
\Omega_{z}(-\frac{1}{2}, \vec n) = - cos\theta.
\end{equation}

The tomographic symbols of the operator $\sigma_{+}$ is obtained from the product of three matrices
\begin{equation}\label{F6}
\begin{aligned}
\Omega_{+}(+\frac{1}{2}, \vec n) =
\left( u \sigma_{+}
u^{+}\right)_{11}
\end{aligned},
\begin{aligned}
\Omega_{+}(-\frac{1}{2}, \vec n) =
\left( u \sigma_{+}
u^{+}\right)_{22}.
\end{aligned}
\end{equation}

Straightforward calculation yields
\begin{equation}\label{G6}
\Omega_{+}(+\frac{1}{2}, \vec n) = \frac{1}{2}\sin\theta e^{\iota
\psi}, \Omega_{+}(-\frac{1}{2}, \vec n) = - \frac{1}{2}\sin\theta
e^{\iota \psi}.
\end{equation}

Using the obtained expression we can construct the tomographic symbols for all
the cases of one, two, three, etc. fermions. For one fermion the tomographic
symbols $\Omega_{a}(m, \vec n)$, $\Omega_{a^{+}}(m, \vec n)$ and
$\Omega_{1}(m, \vec n)$ read
\begin{eqnarray}\label{H6}
\begin{aligned}
\Omega_{a}(m, \vec n) = \Omega_{+}(m, \vec n) = \left\{
\begin{array} {cc}
\frac{1}{2} \sin\theta e^{\iota \psi}, m = +\frac{1}{2} \\
-\frac{1}{2} \sin\theta e^{\iota \psi}, m = -\frac{1}{2}
\end{array} \right.
\end{aligned}
\end{eqnarray}
\begin{eqnarray}\label{I6}
\begin{aligned}
\Omega_{a^{+}}(m, \vec n) = (\Omega_{a})^{*}(m, \vec n) = \left\{
\begin{array} {cc}
\frac{1}{2}\sin\theta e^{-\iota \psi}, m = +\frac{1}{2} \\
-\frac{1}{2}\sin\theta e^{-\iota \psi}, m = -\frac{1}{2}
\end{array} \right.
\end{aligned}
\end{eqnarray}
\begin{eqnarray}\label{J6}
\begin{aligned}
\Omega_{1}(m, \vec n) = \left\{
\begin{array} {cc}
1, m = +\frac{1}{2} \\
1, m = -\frac{1}{2}
\end{array} \right.
\end{aligned}
\end{eqnarray}

For two fermions we introduce the notations $\vec n_j = (sin{\theta}_jcos{\psi}_j, sin{\theta}_jsin{\psi}_j,cos{\theta}_j)$,
$a_j, {a_j}^{+}$; $j=1,2$ as well as $m_j = \pm \frac{1}{2}$.

Thus we get the tomographic symbols ($j=1,2$)
\begin{eqnarray}\label{K6}
\begin{aligned}
\Omega_{+}^{(j)}(m_j, \vec n_j) = \left\{
\begin{array} {cc}
\frac{1}{2}sin{\theta}_je^{\iota {\psi}_j}, m_j = +\frac{1}{2} \\
-\frac{1}{2}sin{\theta}_je^{\iota {\psi}_j}, m_j = -\frac{1}{2}
\end{array} \right .
\end{aligned}
\end{eqnarray}

\begin{equation}\label{L6}
\Omega_{-}^{(j)}(m_j, \vec n_j) = \Omega_{+}^{(j)*}(m_j, \vec n_j).
\end{equation}

\begin{equation}\label{M6}
\Omega_{1}^{(j)}(m_j, \vec n_j) = 1, m_j = \pm \frac{1}{2}.
\end{equation}
\begin{eqnarray}\label{N6}
\begin{aligned}
\Omega_{z}{(j)}(m_j, \vec n_j) = \left\{
\begin{array} {cc}
cos{\theta}_j, m_j = +\frac{1}{2} \\
-cos{\theta}_j, m_j = -\frac{1}{2}
\end{array} \right .
\end{aligned}
\end{eqnarray}
 The tomographic symbols of creation and annihilation operators for two fermions
read
\begin{eqnarray}\label{O6}
\begin{aligned}
\Omega_{a_1}(m_1, \vec n_1, m_2, \vec n_2) =
\Omega_{1}^{(1)}(m_1, \vec n_1)\Omega_{+}{(2)}(m_2, \vec n_2)
= \left\{
\begin{array} {cc}
\frac{1}{2}sin{\theta}_2e^{\iota {\psi}_2}, m_1 = +\frac{1}{2}, m_2 = +\frac{1}{2} \\
-\frac{1}{2}sin{\theta}_2e^{\iota {\psi}_2}, m_1 = +\frac{1}{2}, m_2 = -\frac{1}{2} \\
\frac{1}{2}sin{\theta}_2e^{\iota {\psi}_2}, m_1 = -\frac{1}{2}, m_2 = +\frac{1}{2} \\
-\frac{1}{2}sin{\theta}_2e^{\iota {\psi}_2}, m_1 = -\frac{1}{2},
m_2 = -\frac{1}{2}
\end{array} \right .
\end{aligned}
\end{eqnarray}
\begin{eqnarray}\label{P6}
\begin{aligned}
\Omega_{a_2}(m_1, \vec n_1, m_2, \vec n_2) =
\Omega_{+}{(1)}(m_1, \vec n_1)\Omega_{1}{(2)}(m_2, \vec n_2)
= \left\{
\begin{array} {cc}
\frac{1}{2}sin{\theta}_1e^{\iota {\psi}_1}, m_1 = +\frac{1}{2}, m_2 = +\frac{1}{2} \\
-\frac{1}{2}sin{\theta}_1e^{\iota {\psi}_1}, m_1 = +\frac{1}{2}, m_2 = -\frac{1}{2} \\
\frac{1}{2}sin{\theta}_1e^{\iota {\psi}_1}, m_1 = -\frac{1}{2}, m_2 = +\frac{1}{2} \\
-\frac{1}{2}sin{\theta}_1e^{\iota {\psi}_1}, m_1 = -\frac{1}{2},
m_2 = -\frac{1}{2}
\end{array} \right  .
\end{aligned}
\end{eqnarray}
For creation operators one has
\begin{eqnarray}\label{Q6}
\begin{aligned}
\Omega_{{a_1}^{+}}(m_1, \vec n_1, m_2, \vec n_2) =
{\Omega_{a_1}}^{*}(m_1, \vec n_1, m_2, \vec n_2) , \\
\Omega_{{a_2}^{+}}(m_1, \vec n_1, m_2, \vec n_2) =
{\Omega_{a_2}}^{*}(m_1, \vec n_1, m_2, \vec n_2) .
\end{aligned}
\end{eqnarray}

The formulas for spin-tomographic symbols of creation and annihilation operators
in case of $N$ fermions can be given in the following form
\begin{eqnarray}\label{R6}
\begin{aligned}
 \Omega_{a_j}(m_1, \vec n_1, m_2, \vec n_2, \cdots m_N, \vec n_N)= \\
 \hspace{20mm} = \Omega_{+}^{(N-j+1)}(m_{N-j+1}, \vec n_{N-j+1})\prod_{k=N-j+2}^{N} \Omega_{z}^{k}(m_k, \vec n_k)
\end{aligned}
\end{eqnarray}

The formulas for factors $\Omega_{+}^{s}(m_s, \vec n_s)$ where
$s=N-j+1$ and for $f_{z}^{k}(m_k, \vec n_k)$ are given by (\ref{J6})
and (\ref{M6}). The symbol of creation Fermi operator ${a_j}^{+}$
is given by (\ref{R6}) with the replacement $\Omega_{+}^{(N-j+1)}
\longrightarrow \Omega_{-}^{(N-j+1)}$. We give tomogram of $\mid
vac\rangle$ for $N=2$
\begin{eqnarray}\label{S6}
\begin{aligned}
\omega_{vac}(m_1, \vec n_1, m_2, \vec n_2) = \left\{
\begin{array} {cc}
\cos^2\frac{\theta_1}2\cos^2\frac{\theta_2}2, m_1 = +\frac12, m_2 = +\frac12 \\
\cos^2\frac{\theta_1}2\sin^2\frac{\theta_2}2, m_1 = +\frac12, m_2 = -\frac12 \\
\sin^2\frac{\theta_1}2\cos^2\frac{\theta_2}2, m_1 = -\frac12, m_2 = +\frac12 \\
\sin^2\frac{\theta_1}2\sin^2\frac{\theta_2}2, m_1 =
-\frac{1}{2}, m_2 = -\frac{1}{2}
\end{array} \right .
\end{aligned}
\end{eqnarray}

The star-product kernel of the tomographic symbols for $N$ fermions is the product of kernels
for each of the fermions which is given by formula (\ref{N4}). It follows from the fact that the
dequantizers and quantizers for several fermions are given as the tensor product of
quantizers and dequantizers for each of fermions.

\section {Conclusion}
To conclude we point out the main results of our work. The known
anticommutation relations for fermion field operators were
presented in explicit matrix realization. Using the matrix
realization we mapped the fermion field operators onto spin
tomographic symbols. The density matrix of fermion vacuum state of
N fermions was used to construct the tomographic probability
representation of this state and the tomogram of the fermion
vacuum is calculated in explicit form. The creation and
annihilation fermion operators and their multiplication are given
in the tomographic framework by using explicit tomographic
star-product kernel.

\section*{Acknowledgement}
M.A.Man'ko, V.I.Man'ko,Yu.P.Timofeev and S.D.Zakharov
are grateful to the Russian Foundation for Basic Research for
partial support under Project No.~08-03-90300. Nguyen Hung Son and
Nguyen Cong Thanh would like to thank the Vietnam Academy of
Science and Technology for the financial support and the Lebedev
Physical Institute for the kind hospitality.


\begin{thebibliography}{99}

\bibitem{Dirac}
P.A.M.Dirac, {\it The Principle of Quantum Mechanis}, 4th edition
Pergamon, Oxford (1958).

\bibitem {Scr}
E.Schrodinger, {\it Ann.Phys.} (Leipzig), {\bf 79}, 489 (1926).
\bibitem {Landau}
L.D.Landau, {\it Z.Phys.}, {\bf 45}, 430 (1927).
\bibitem {van Neumann}
J. von Neumann, {\it Mathematische Grundlagen der Quantummechanyk}, Springer, Berlin (1932).
\bibitem{Bog}
N.N.Bogolubov and D.V.Shirkov, {\it Introduction to the theory of
quantum field} (in Russian), Nauka, Moscow (1976).

\bibitem{NVH}
Nguyen Van Hieu, {\it The method of second quantization}(in Russian),
Energoatomizdat, Moscow (1984).

\bibitem{Berezin}
F.A.Berezin, {\it Sov.Phys.Izv.Akad.Nauk} {\bf 8}, 1109 (1974).


\bibitem{Manc 1996}
S. Mancini, V.I. Man'ko and P.Tombesi, {\it Phys. Lett. A}, {\bf 213}, 1
(1996).

\bibitem{Vent}
A. Ibort, V.I. Man'ko, G. Marmo, A. Simoni, and F. Ventriglia {\it Phus.Scr.} {\bf 79}, 065013 (2009).


\bibitem{Rosa}
V. I. Man'ko, L. Rosa, and P. Vitale, {\it Phys. Lett. B}, {\bf 439}, 328
(1998).

\bibitem {Stern1}
V.I.Man'ko, G.Marmo and C.Stornaiolo, {\it Gen.Relativ.Gravit.}, {\bf 37}, 99 (2005).

\bibitem {Stern2}
V.I.Man'ko, G.Marmo and C.Stornaiolo, {\it Gen.Relativ.Gravit.}, {\bf 37}, 2003 (2005).

\bibitem {Son}
M.A.Man'ko, V.I.Man'ko, Nguyen Hung Son, Nguyen Cong Thanh, Yu.P.Timofeev and
S.D.Zakharov, {\it J.Russ.Laser.Res.}, {\bf 30}, 1 (2009).

\bibitem{Mar}
O.V. Man'ko, V.I. Man'ko and G. Marmo, {\it J. Phys. A: Math.Gen.}, {\bf 35}, 699
(2002).

\bibitem {D}
V.V.Dodonov and V.I.Man'ko, {\it Phys.Lett. A}, {\bf 229}, 335 (1997).

\bibitem {O}
O.V. Man'ko and V.I.Man'ko, {\it J.Exp.Theor.Phys.}, {\bf 85}, 430 (1997).

\bibitem {An}
V.A.Andreev, V.I.Man'ko, O.V.Man'ko and E.V.Shchukin, {\it Theor.Math.Phys.}, {\bf 146}, 172 (2006).

\bibitem {Dav}
A.S.Davydov, {\it Quantum Mechanics} (in Russian), Nauka, Moscow (1965).

\bibitem {Stratonov}
R.L.Stratonovich, {\it J.Exp.Theor.Phys.}, {\bf 31}, 1012 (1956).

\bibitem {Fronsdal}
F.Bayern, M.Flato, C.Fronsdal et al., {\it Lett.Math.Phys.}, {\bf 1}, 251 (1975).

\bibitem {Fil}
S.N.Filippov and V.I.Man'ko, {\it J.Russ.Laser.Res.}, {\bf 30}, 82 (2009).

\end{thebibliography}
\end{document}